\documentclass[12pt]{article}
\usepackage{epsfig}
\usepackage{graphics}
\usepackage{times}
\begin{document}

\title{Why Does the Solar Corona Abnormally Rotate Faster Than the Photosphere?}
\author {K. J. LI$^{1, 3, 4}$, J. C. XU$^{1, 3, 4}$, Z. Q. Yin$^{4}$, W.  FENG$^{2}$     \\
\footnotesize{$^{1}$Yunnan Observatories, Chinese Academy of Sciences, Kunming 650011, China}\\
\footnotesize{$^{2}$Research Center of Analysis and Measurement, Kunming University of Science and Technology, Kunming 650093, China}\\
\footnotesize{$^{3}$Center for Astronomical Mega-Science, Chinese Academy of Sciences, Beijing 100012, China} \\
\footnotesize{$^{4}$Key Laboratory of Solar Activity, National Astronomical Observatories, CAS, Beijing 100012, China} \\
}

\date{}
\baselineskip24pt
\maketitle

\begin{abstract}
Coronal heating is a big question for modern astronomy.
Daily measurement of 985 solar spectral irradiances (SSIs) at the spectral intervals 1-39 nm and 116-2416 nm during March 1 2003 to October 28 2017  is utilized to investigate
characteristics of solar rotation in the solar atmosphere by means of the Lomb \,-\, Scargle periodogram method to calculate their power spectra.
The rotation period of coronal plasma is obtained to be 26.3 days, and that of the solar atmosphere at the bottom of the photosphere modulated by magnetic structures is 27.5 days.
Here we report for the first time that unexpectedly the coronal atmosphere is found to rotate faster than the underlying photosphere.
When time series of SSIs are divided into different cycles, and the ascending and descending periods of a solar cycle, rotation rate in the corona is also found to be larger than that in
the photosphere, and
this actually gives hidden evidence: it is small-scale magnetic activity that heats the corona. \\
{\bf keywords}  Sun: rotation-- Sun: corona-- Sun: atmosphere
\end{abstract}

\section{Introduction}
When energy flows from the interior of the Sun outward into the solar atmosphere,
why is the Sun's  outer atmosphere, the corona
much hotter than the inner atmosphere, the underlying chromosphere and photosphere?  This is the long-standing problem of the coronal heating, which is one of the eight key mysteries in modern astronomy (Kerr 2012).
For about 80 years since the discovery of extremely hot corona around the late 1930s (Grotian 1939; Edlen 1945), people have worked hard on addressing this issue, and great advances have been achieved in observation and theoretical studies (Parnell 2012; Amari et al. 2015; Arregui 2015; Cargill et al 2015; De Moortel $\&$ Browning 2015; Jess et al 2015; Klimchuk 2015; Longcope $\&$ Tarr 2015; Peter 2015; Schmelz $\&$ Winebarger 2015;  Velli et al 2015; Wilmot-Smith 2015). Especially during recent decades,
high-resolution observations of solar super-fine structures indicate that small spicules, minor hot jets along small-scale magnetic channels from the low atmosphere upwards to the corona, petty  tornados and  cyclones, and small explosive phenomena such as mini-filament eruptions and micro- and nano-flares, all of these small-scale magnetic activities contribute greatly to coronal heating (De Pontieu et al 2011; Zhang $\&$ Liu 2011; Parnell 2012; Klimchuk 2015; Peter 2015; Schmelz $\&$ Winebarger 2015;  Henriques et al 2016; De Pontieu et al 2018; Li et al 2018). Also contributions of MHD waves to heating the corona have been observationally illustrated (van Ballegooijen et al 2011; Jess et al 2015; Kubo et al 2016; Morton et al 2016; Soler et al 2017; Morgan $\&$ Hutton 2018).
Meanwhile with the progress of observation studies, two groups of theoretical models, magnetic reconnection models and magnetohydrodynamic wave models have traditionally attempted to explain coronal heating,  but so far no models can address the key mystery perfectly (van Ballegooijen et al 2011; Arregui 2015; Cargill et al 2015; Peter 2015; Velli et al 2015;  Wilmot-Smith 2015). Maybe we do not need to intentionally take to heart  such the classical dichotomy, because
waves and reconnections may interact on each other (De Moortel $\&$ Browning 2015; Velli et al 2015). Additionally, statistical studies may look at coronal heating from a comprehensive perspective. Li et al (2018) found that the long-term variation of the heated corona, which is represented by  coronal spectral irradiances, and that of small-scale magnetic activity  are in lockstep, indicating that the corona should statistically be effectively heated by small-scale magnetic activity. Observation and theoretical model studies  through heating channels and modes, and statistical studies by means of heating effect (performance of the heated corona), both point the finger of the coronal heating firmly at small-scale magnetic activity.

Solar rotation is one of elementary characteristics of the Sun, and it is the most prominent period except the  Schwabe cycle of about 11 years (Howard 1984; Vats et al 1998; Vats et al. 2001; Chandra, Vats,  $\&$  Iyer 2009; W$\ddot{o}$hl et al 2009; Zaatri et al 2009; Chandra $\&$ Vats 2011; Xie,  Shi,  $\&$ Xu 2012£»Li et al 2013).
Many solar activity events of long life in the solar atmosphere and interior present the rotation period of about 27 days,
and investigations of solar rotation and further its relationship with solar activity are one of important observational studies in solar physics and have been paid attentions to in contemporary observations of  both high temporal and spatial resolution (Howard 1984; Vats et al 1998; W$\ddot{o}$hl et al 2009; Zaatri et al 2009£» Chandra, Vats,  $\&$  Iyer 2010; Vats $\&$   Chandra 2011; Li et al. 2014£» Sudar et al. 2016; Bhatt et al. 2017).
Actual measurements of  solar (differential) rotation can provide a constraint through observations on solar dynamo  models, because  differential rotation plays an important role in the translation of the toroidal magnetic field from the polar magnetic field (Babcock 1963; Duvall 1984; Thompson et al 1996; Schou et al 1998; Howe 2009). There are two main ways to probe rotation signals, one is  observational measurement, including tracer and spectrum measurements, and the other is  period analysis of time series of solar activities (Howard 1984; Vats et al 1998; W$\ddot{o}$hl et al 2009; Zaatri et al 2009).  Energy and even some matter of the solar  atmosphere come from the solid sun, namely the solar body, and the rotation of the former is driven by the rotation of the latter, therefore the rotation of the former should not be faster than the rotation of the latter, and the rotation of the solar upper atmosphere should not be faster the rotation of the solar bottom atmosphere.
In this study, 985 solar spectral irradiances are used to investigate rotation characteristics of the entire solar atmosphere, from the low photosphere to the high corona, and the coronal is found to rotate abnormally quicker than the underlying photosphere atmosphere.

\section{Rotation period of the solar layered atmosphere}
\subsection{Data  and method}
Daily solar spectral irradiances (SSIs)
at 985 disjunct bands of spectrum intervals 1-39 nm and 116-2416 nm are measured by the SORCE satellite during March 1 2003 (the year of 2003.162) to October 28 2017 (the year of 2017.823). They are available from http://lasp.colorado.edu/home/sorce/data/. Here Figure 1 shows 6 seleted bands of SSIs as  samples, and a movie for all bands was once given by Li et al (2018).

\begin{figure*}
\hskip  5.0 cm
\centerline{\includegraphics[width=1.1\textwidth]{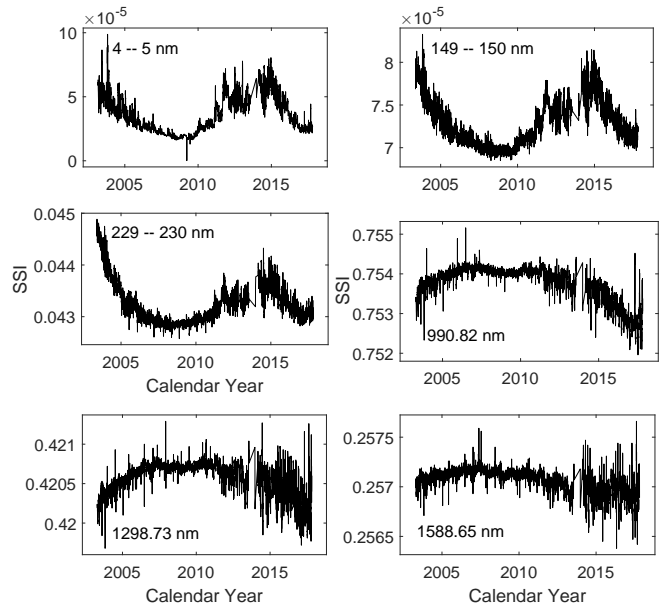}}
\hskip  5.0 cm
\caption{Daily solar spectral irradiance
at 6 spectral bands  measured by the SORCE satellite during March 1 2003 to October 28 2017. The band of a SSI is shown at the upper  left corner of each panel.
}\label{}
\end{figure*}

Figure 2 shows the number of data days and that of no-data days for each SSI, and the total time span of a SSI series is 5356 days. The 985 time series of SSI are all uneven, and especially those SSIs whose wavelengths are longer than 1600 nm are more obviously uneven than the SSIs whose wavelengths are shorter than 1600 nm. For the latter,
the number of no-data days is basically about 370 days, accounting for about $7\%$ of the total.
The Lomb \,-\, Scargle periodogram method (Lomb 1976; Scargle 1982; Horne $\&$ Baliunas 1986) can ease the puzzle of missing data (Deng et al 2013; Xiang $\&$ Qu 2016; Xu $\&$ Gao 2016; Xie et al 2018), and thus it is utilized here to investigate  periodicity of each SSI.

\begin{figure*}
\hskip  5.0 cm
\centerline{\includegraphics[width=1.0\textwidth]{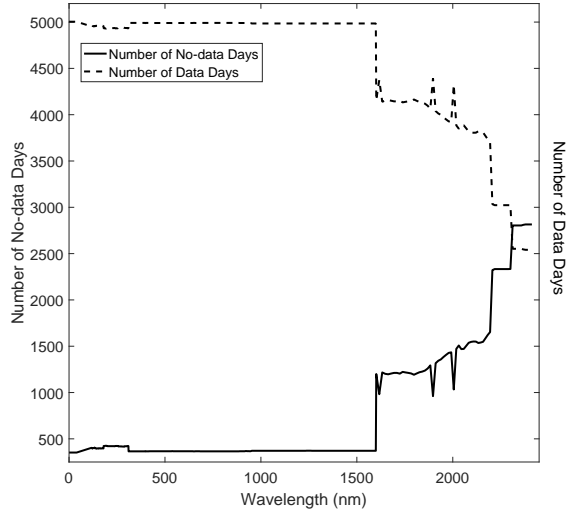}}
\caption{The number of data/no-data days for SSIs at all 985 bands (dashed/solid line).
}\label{}
\end{figure*}
\begin{figure*}
\hskip  5.0 cm
\centerline{\includegraphics[width=1.0\textwidth]{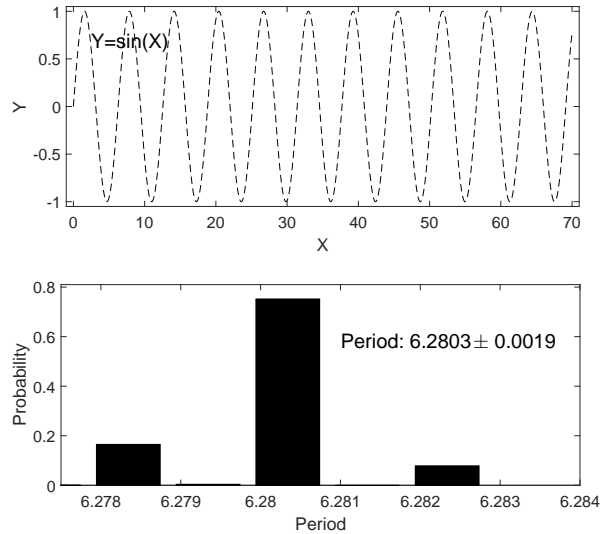}}
\caption{Top panel: The periodical function $Y=sin(X)$,  where X is evenly taken discrete values  from 0 to 70, and the distance between two adjacent values is always 0.02.
Bottom panel: Probability distribution of 2000 period values of the function, which are determined by means of the Lomb \,-\, Scargle periodogram method (for details, please see the text).  }\label{}
\end{figure*}

Here first of all, the periodical function, $Y=sin(X)$ as an example is utilized to test the reliability of the Lomb \,-\, Scargle method. The function is  shown in Figure 3,  where $X$ is evenly taken to go from 0 to 70, with a constant spacing, 0.02, and thus the total  number of data points is 3501. Each time, $20\%$ of the data points are randomly removed, and the rest $80\%$  are utilized to calculate periodicity of the function by means of the Lomb \,-\, Scargle method. Such a process is repeated 2000 times, and 2000 period values are obtained, which are shown in Figure 3. The maximum of these period values is 6.2823, the minimum is 6.2753, and the average is $6.2803\pm 0.0019$. The real period of the function is known to be $2\pi$, so  the periodogram method is feasible. 

\subsection{Rotation period for the entire considered time interval}
As samples, the power spectra of the 6 selected bands of SSI are shown in Figure 4, and also given are their corresponding statistical test lines at the $0.1\%$ significance level.
The power spectra of all 985 SSIs are made into an animation, which is given here as an attachment.
This study aims to investigate rotation characteristics of SSIs, so the figure only shows the power spectra at a period range around the rotation-period scale of about 27 days, from 22 to 33 days.
As the figure shows, sometimes just one peak around the rotation period scale is statistically significant at the $0.1\%$ level, and such a case is found to occur for the SSIs at the bands of 1-40 nm and 980-1600 nm, but sometimes peaks of statistical significance are clustered around the rotation period scale.
Here  for a special SSI, the period corresponding to the maximum power in the period-scale range of 22 to 33 days is regarded as the rotation period of the SSI.
A SSI may give a rotation-period value, and
Figure 5 shows the determined synodical period of solar rotation for each SSI while its statistical significance is not considered, and in the followings, such similar figures will be given to just comprehensively show potential periods. Further, Figure 6 shows those determined periods of solar rotation which are
statistically significant at the $0.1\%$ level, and in the followings, further investigations will be carried on  according to such similar figures. As these two figures display,   synodical rotation periods can be divided into four parts, according to distribution characteristics of rotation period varying with SSI.
For Part 1 {\bf (marked as P1 in Figures 5 and 6)},  wavelength ($L$) of SSI is shorter than 264 nm, and rotation period is basically a constant. The averaged {\bf (synodic)} rotation period over the whole part is $26.332\pm 0.120$ (days) in Figure 5, and $26.327\pm0.129$ (days) in Figure 6. The relation between synodic rotation period ($T_{syn}$) and sidereal rotation period ($T_{sid}$) is that $T_{sid}=365.256\times T_{syn}/(365.256+T_{syn})$, and their corresponding errors are related as that: $\Delta T_{sid}=365.256^{2}/(365.256+T_{syn})^{2}\times  \Delta T_{syn}$. Therefore correspondingly, the averaged sidereal rotation period
is $24.561\pm 0.104$ and $24.557\pm 0.112$ in days, respectively.
For this part there are 189 points in Figure 5, and 161 points in Figure 6.
For Part 2 (marked as P2 in Figures 5 and 6), 264 nm $< L <$ 948 nm, rotation period obviously varies with wavelength  $L$.
For Part 3 (marked as P3 in Figures 5 and 6), 952 nm $< L <$ 1600 nm, rotation period returns to be basically a constant, similar  to Part 1. The averaged  (synodical) rotation period over the whole part is $27.551\pm 0.057$ (days) in Figure 5 and $27.545\pm0.055$ (days) in Figure 6, and correspondingly, the averaged sidereal rotation period
is $25.619\pm 0.049$ and $25.613\pm 0.048$ in days, respectively.
For this part there are 119 points in Figure 5, and 111 points in Figure 6.
For Part 4 (marked as P4 in Figures 5 and 6), $L >$ 1601 nm, rotation period clearly varies with wavelength  again, similar to Part 2.
\begin{figure*}
\hskip  5.0 cm
\centerline{\includegraphics[width=1.0\textwidth]{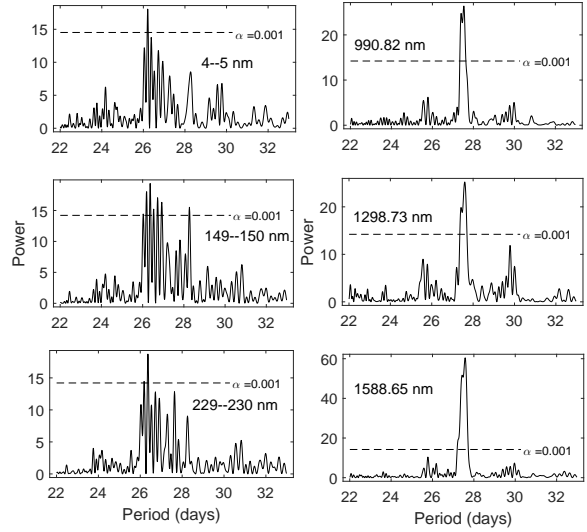}}
\caption{Power spectra of the sample SSIs at the 6 spectral bands which are shown in figure 1.
The x-axis shows period scale in days, and the y-axis, power.
The dashed lines indicate the $0.01\%$ significance level.  An animation of all periodograms of SSIs at the 985 spectral bands is available as an attachment.
}\label{}
\end{figure*}
\begin{figure*}
\hskip  5.0 cm
\centerline{\includegraphics[width=1.0\textwidth]{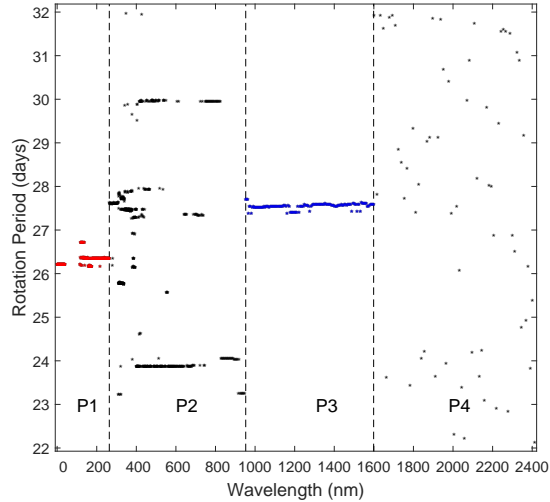}}
\caption{Synodical rotation period in days (asterisks) corresponding to the maximum power among a band of power spectra of a SSI around rotation period scale of about 27 days, from 22 to 33 days, when its statistical significance is not considered. A time series of SSI gives a rotation period, and 985 rotation-period values of 985 SSIs can be divided into 4 parts, marked as P1, P2, P3, and P4. Rotation periods hardly change with SSI in P1 and P3, which are correspondingly marked by red and blue color.
}\label{}
\end{figure*}
\begin{figure*}
\hskip  5.0 cm
\centerline{\includegraphics[width=1.0\textwidth]{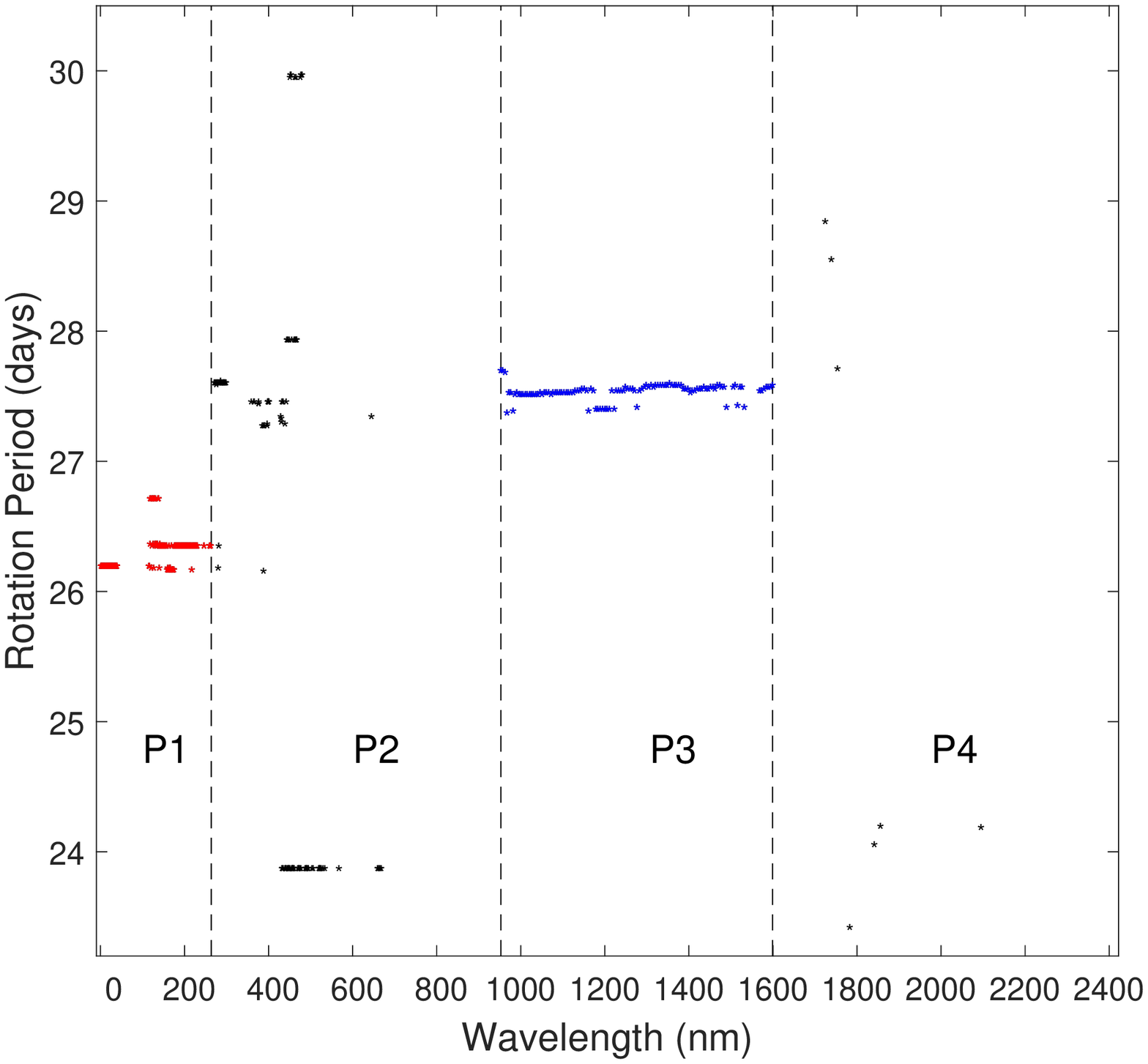}}
\caption{The same as Figure 5, but shown are just those periods  which are statistically significant at the $0.1\%$ level.
}\label{}
\end{figure*}

In Figure 6, the difference of the mean rotation period between Parts 1 and 3 with the $99\%$ probability should be located at the interval $27.545-26.327\pm Z_{0.005}\sqrt{0.129^{2}/161+0.055^{2}/111}$ (days), namely $1.218\pm 0.029$ (days), where the tabulated critical value of the normal distribution, $Z_{0.005}=2.576$. Therefore, the mean rotation period is significantly longer in Part 3 than that in Part 1. By the way in Figure 5  as well, the mean rotation period is found to be significantly longer in Part 3.
Actually in Figures 5 and 6, each of the individual rotation-period  values  in Part 3 is larger than that in Part 1, also indicating that rotation period should be in general  significantly longer in Part 3 than that in Part 1.

\subsection{Rotation period for the descending time of cycle 23}
The end epoch of cycle 23 is the year of 2008.958, and the Lomb \,-\, Scargle periodogram method is used to calculate individual power spectra of 985 SSI in the descending period of cycle 23 (from March 1 2013 to the year of 2008.958). As  samples, Figure 7 shows the power spectra of the 6 selected bands of SSI  and their corresponding statistical test lines at the $0.1\%$ significance level. The same method mentioned in the last subsection is used here and will be used in the followings to determine rotation periods.
Similarly, Figure 8 shows the determined period of solar rotation for each SSI while its statistical significance is not considered, and Figure 9 displays those determined periods of solar rotation which are statistically significant at the $0.1\%$ level.
These two figures indicate that the distribution of synodical rotation period varying with SSI can be divided into the same four parts as Figures 5 and 6 do, and
for Parts 1 and 3 (P1 and P3), rotation period is basically a constant, much slightly varying with wavelength. For each of these two parts, the averaged rotation period ($\bar{P}$) over the whole part and the corresponding variance ($\sigma$) are given in Table 1.  Given in the table are the number ($N$) of significant rotation periods included in each part, namely the number of data points of each part in Figure 9, and the probability of that one averaged rotation period is longer than the other. The table indicates that the mean rotation period is significantly longer in Part 3 than that in Part 1.  Actually in Figures 8 and 9, each of the individual rotation-period  values  in Part 3 is strikingly larger than that in Part 1, also indicating that rotation period should be in general  significantly longer in Part 3 than that in Part 1.

\begin{figure*}
\hskip  5.0 cm
\centerline{\includegraphics[width=1.1\textwidth]{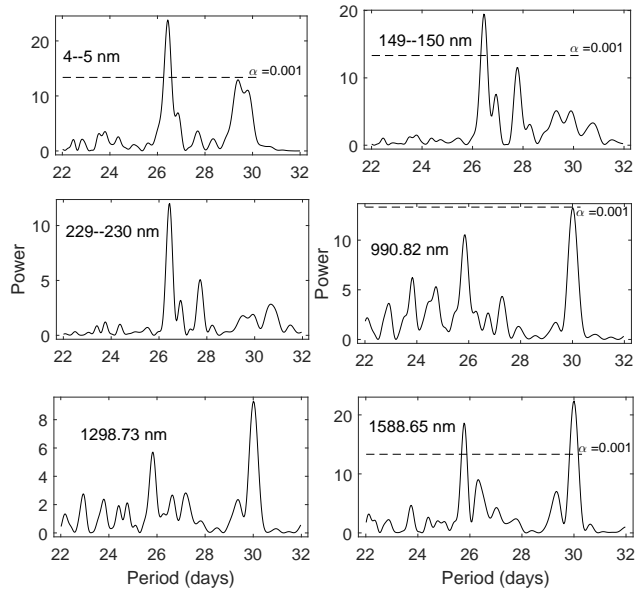}}
\caption{The same as Figure 4, but just SSIs in the descending time of cycle 23 (March 1 2003 to the year of 2008.958)  are used to calculate power spectra.
}\label{}
\end{figure*}
\begin{figure*}
\hskip  5.0 cm
\centerline{\includegraphics[width=1.0\textwidth]{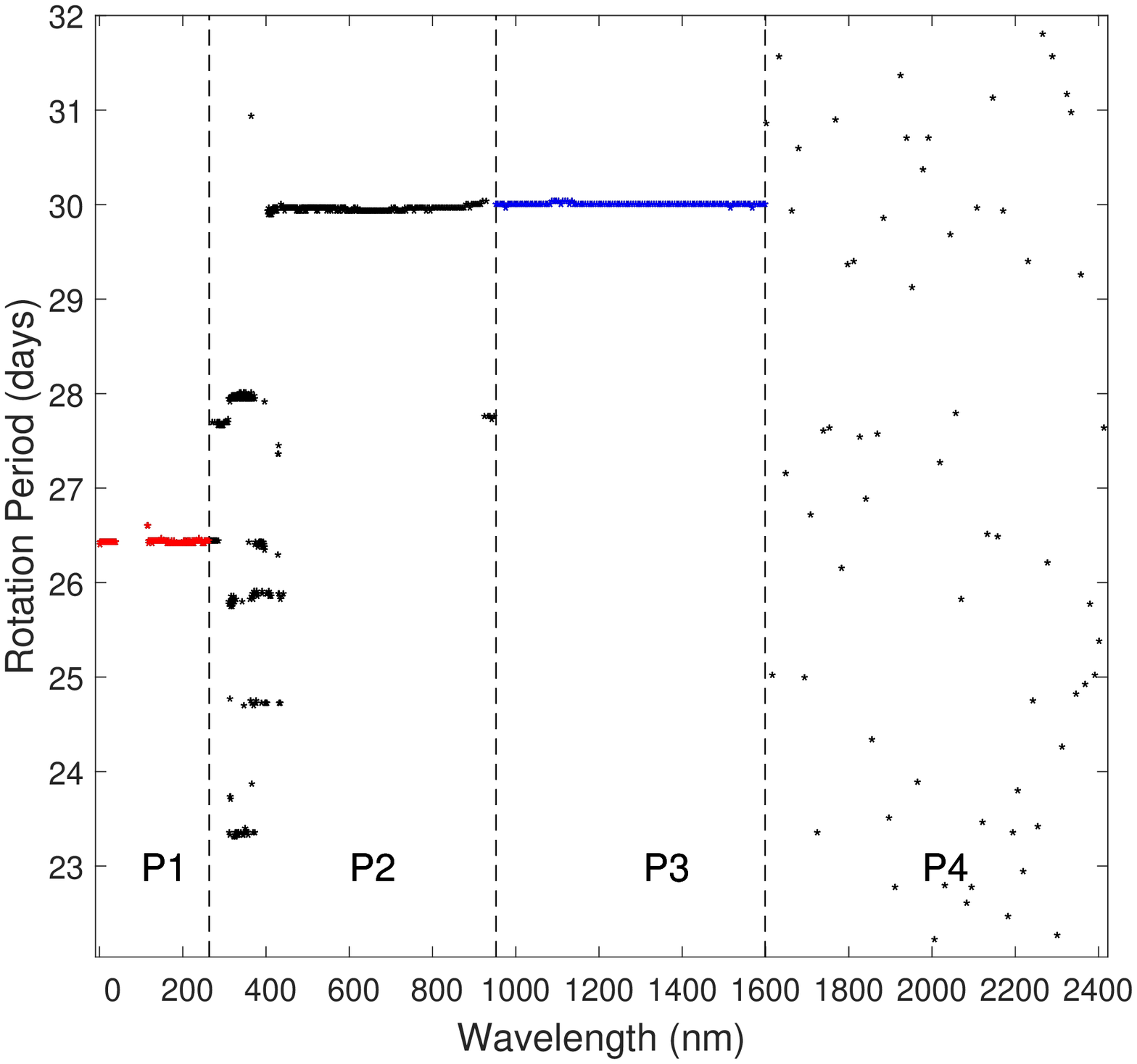}}
\caption{The same as Figure 5, but just SSIs in the descending time of cycle 23  are used to to calculate power spectra, and then to determine rotation periods.
}\label{}
\end{figure*}
\begin{figure*}
\hskip  5.0 cm
\centerline{\includegraphics[width=1.0\textwidth]{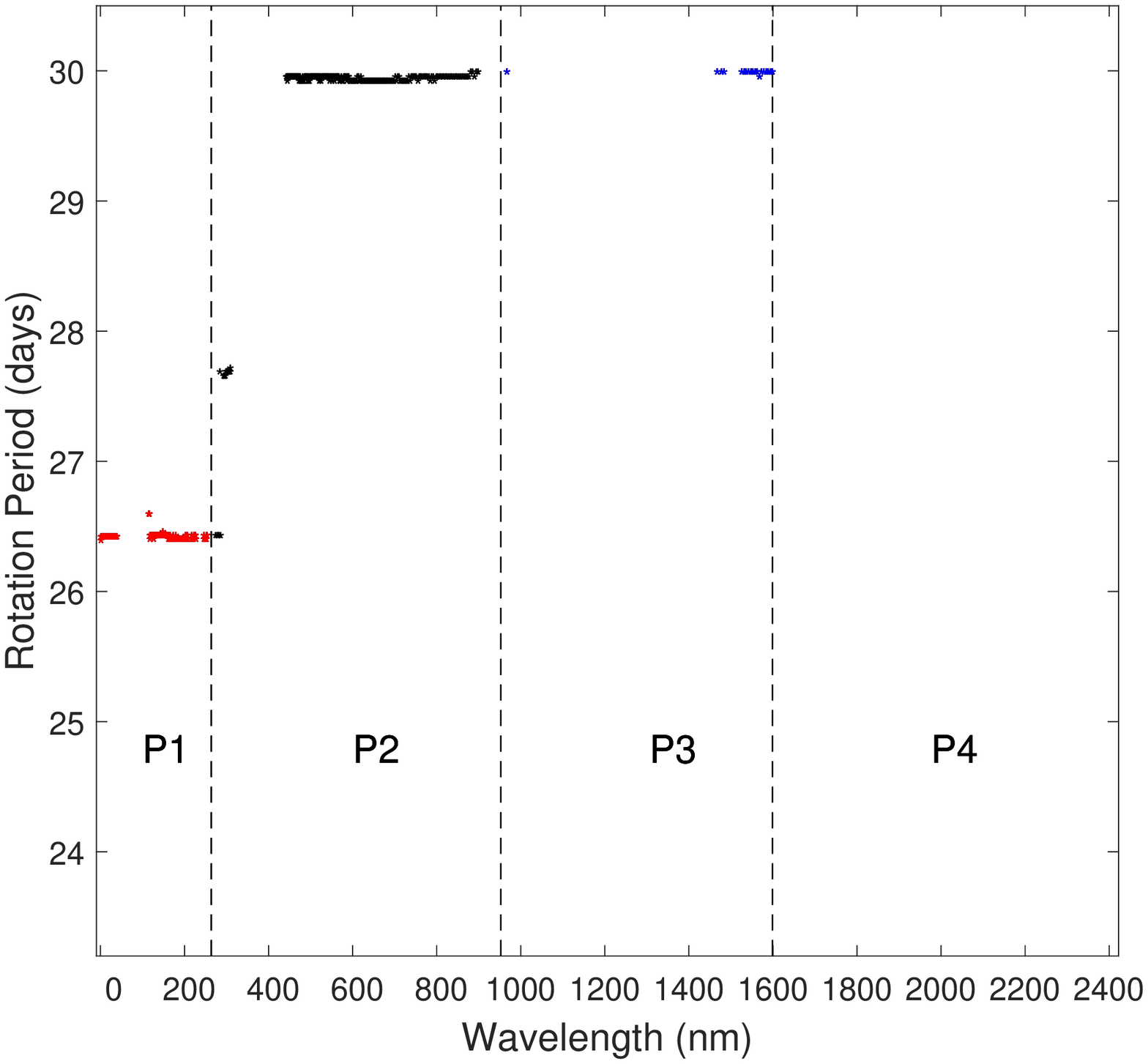}}
\caption{The same as Figure 6, but just SSIs in the descending time of cycle 23  are used to calculate power spectra, and then to determine rotation periods.
}\label{}
\end{figure*}

%
\begin{table}
\leftskip -15 mm
 \caption{The averaged rotation period in days over each of Parts 1 and 3}
  \begin{tabular}{llllllll}
\hline
  Time interval             & \multicolumn{3}{c}{Part 1}                      & \multicolumn{3}{c}{Part 3} &     Probability  \\
                         & $T_{syn}\pm \sigma$ & $N$ & $T_{sid}$         & $T_{syn}\pm \sigma$ & $N$  & $T_{sid}$         &     $> 99\%$     \\
$DT^{1}$ of $SC^{2}$ 23  &  $26.434\pm 0.023$  & 154 & $24.650\pm 0.020$ & $30.004\pm 0.008$   &  19  & $27.726\pm 0.007$ &     $> 99\%$     \\
SC 24                    &  $26.137\pm 0.201$  & 184 & $24.391\pm 0.175$ & $27.498\pm 0.014$   &  119 & $25.573\pm 0.012$ &     $> 99\%$     \\
$AT^{3}$ of SC 24        &  $26.804\pm 0.146$  & 87  & $24.971\pm 0.127$ & $27.549\pm 0.108$   &  119 & $25.617\pm 0.093$ &     $> 99\%$     \\
DT of SC 24              &  $26.265\pm 0.029$  & 181 & $24.503\pm 0.025$ & $27.479\pm 0.024$   &  118 & $25.556\pm 0.021$ &     $> 99\%$     \\
 \hline
 \hline
  \end{tabular}
$^{1}DT:$  Descending time. $^{2}SC:$ Solar cycle. $^{3}AT:$  Ascending time.
\end{table}
%

\subsection{Rotation period for cycle 24}
Cycle 24 starts from December 2008,  and here the same periodogram method is used to calculate individual power spectra of 985 SSIs in  cycle 24 (from  the year of 2008.958 to October. 28, 2017). Figure 10 shows the power spectra of the 6 selected bands of SSI  and their corresponding statistical test lines at the $0.1\%$ significance level.  Then
similarly, Figure 11 shows the determined period of solar rotation for each SSI while its statistical significance is not considered, and Figure 12 displays those determined periods of solar rotation which are statistically significant at the $0.1\%$ level.
These two figures indicate that the distribution of synodical rotation period varying with SSI can be divided into the same four parts as Figures 5 and 6 do, and
for Part 3 (P3), rotation period is basically a constant, much slightly varying with wavelength.
However for Part 1 (P1), rotation period  much slightly varies with wavelength, except 5 of the total 184 wavelength bands in Figure 12. If these 5 rotation values are excluded, the mean rotation of this part should change to be $26.102\pm 0.018$ (days) from the the original value $26.137\pm 0.201$ (days), with variance being obviously reduced.

\begin{figure*}
\hskip  5.0 cm
\centerline{\includegraphics[width=1.1\textwidth]{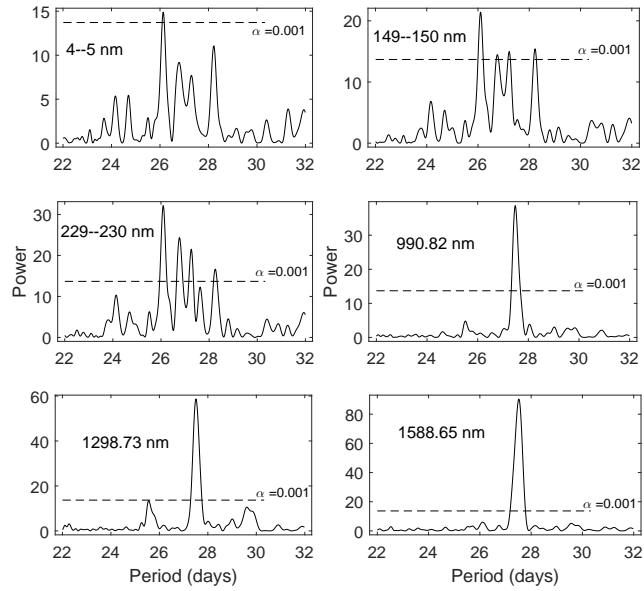}}
\caption{The same as Figure 4, but just SSIs in cycle 24 (the year of 2008.958 to October 28 2017)  are used to calculate power spectra.
}\label{}
\end{figure*}
\begin{figure*}
\hskip  5.0 cm
\centerline{\includegraphics[width=1.0\textwidth]{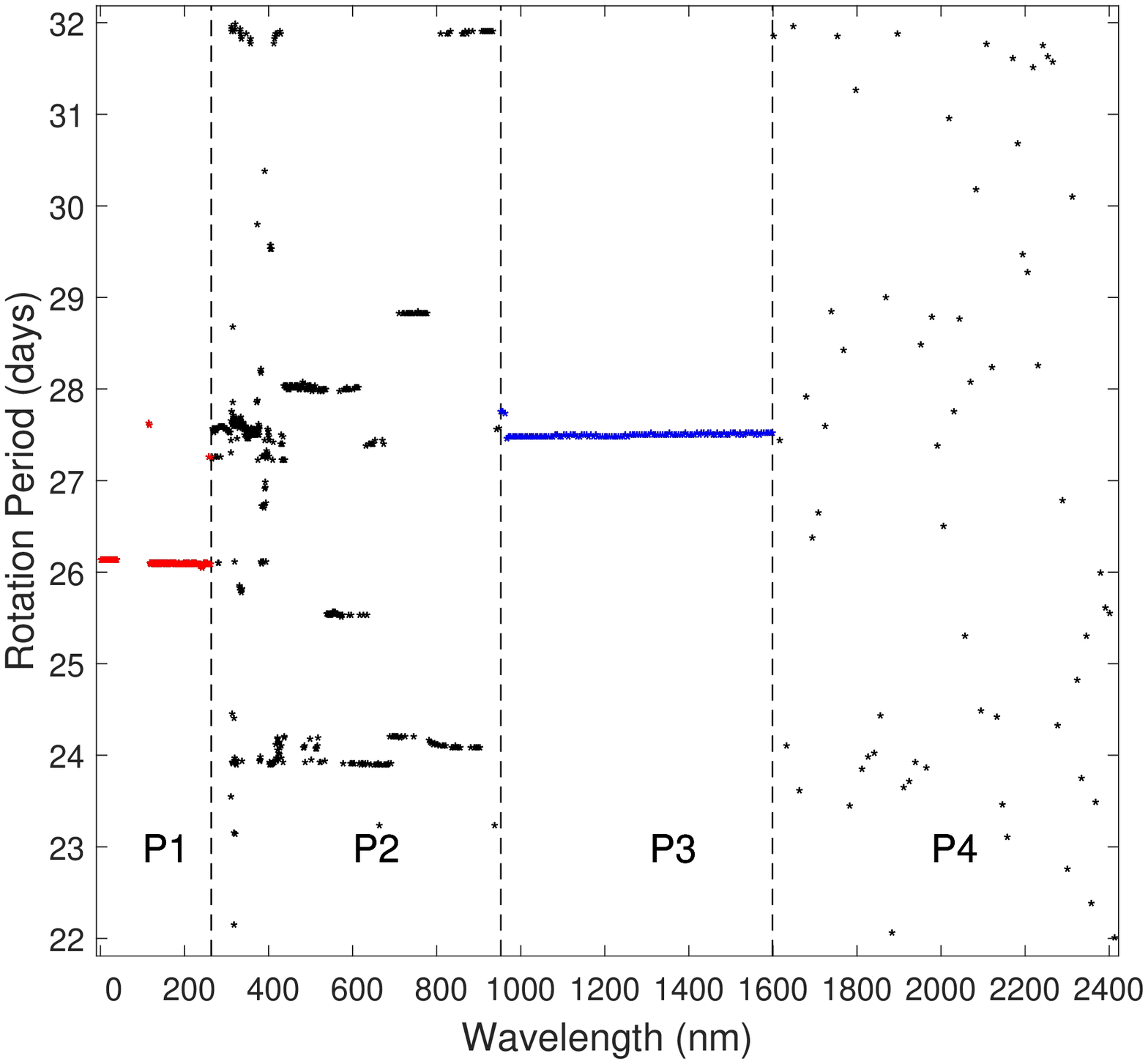}}
\caption{The same as Figure 5, but just SSIs in cycle 24   are used to to calculate power spectra, and then to determine rotation periods.
}\label{}
\end{figure*}
\begin{figure*}
\hskip  5.0 cm
\centerline{\includegraphics[width=1.0\textwidth]{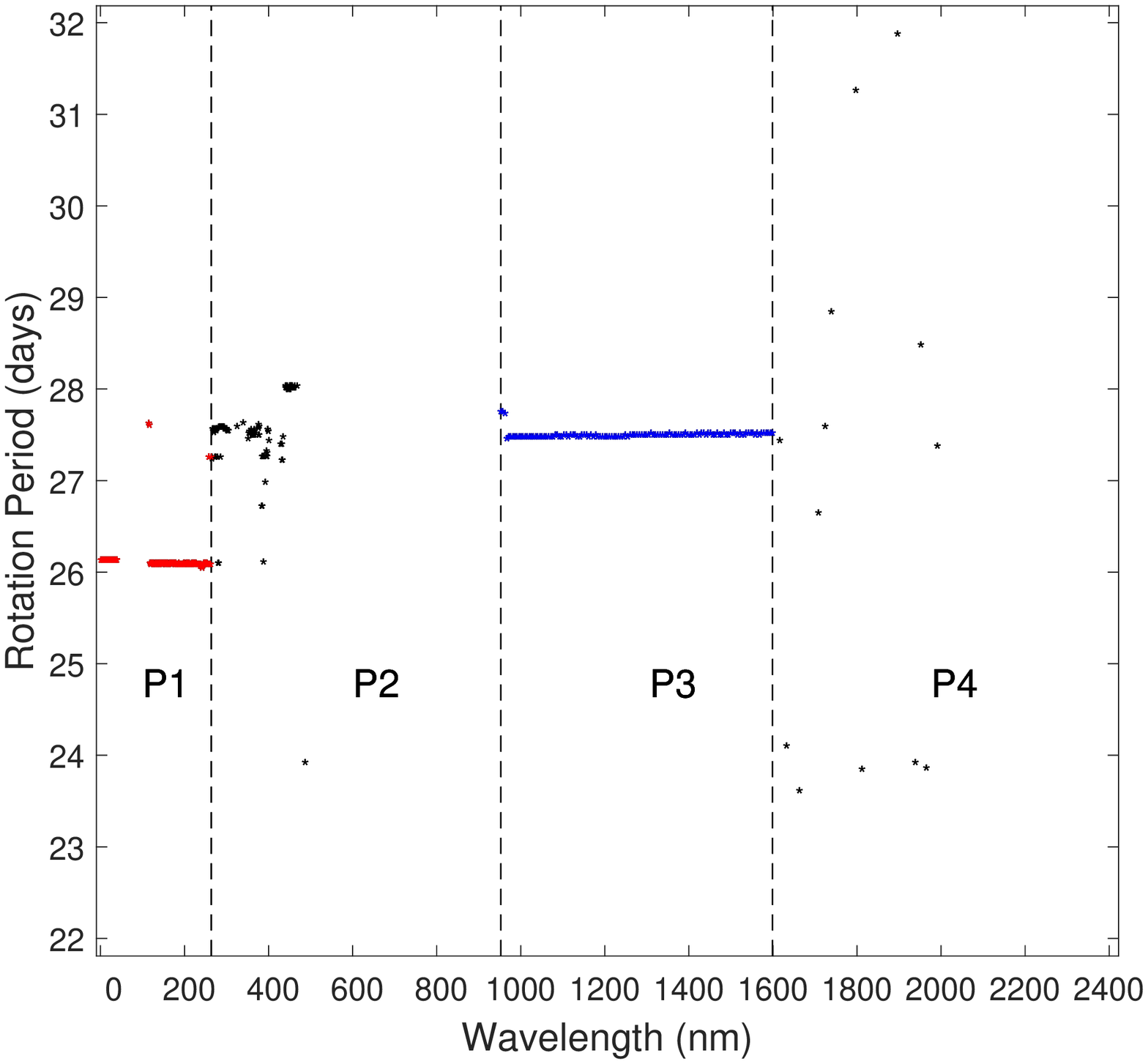}}
\caption{The same as Figure 6, but just SSIs in cycle 24  are used to calculate power spectra, and then to determine rotation periods.
}\label{}
\end{figure*}

Similarly, the relevant values on the obtained results  are given in Table 1.
and it indicates that the mean rotation period is significantly longer in Part 3 than that in Part 1.  Actually in Figures 11 and 12, all rotation-period  values except 5 values  in Part 1 are clearly smaller than those in Part 3, also indicating that rotation period should be in general  significantly longer in Part 3 than that in Part 1.

\subsection{Rotation period for the ascending time of cycle 24}
Cycle 24 peaks in April 2014 (the year of 2014.292),  and 985 SSIs in  the ascending time of cycle 24 (from  the years of 2008.958 to 2014.292)
are used to calculate individual power spectra by means of the same periodogram method. The power spectra of the 6 selected bands of SSI  are  shown in Figure 13 as samples.  Then
similarly, Figure 14 shows the  period of solar rotation for each SSI while its statistical significance is not considered, and Figure 15, those  periods of solar rotation which are statistically significant at the $0.1\%$ level.
The distribution of synodical rotation period varying with SSI can be divided into the same four parts as Figures 5 and 6 do, and
for Parts 1 and 3, rotation period slightly varies with wavelength.
Table 1 gives the relevant values of rotation periods, and
the mean rotation period  in Part 3 is found to be significantly longer than that in Part 1.  Actually in Figures 14 and 15, although rotation periods in Parts 1 and 3 slightly go up and down, all rotation-period  values in Part 3 are distinctly larger than those in Part 1, indicating that rotation rate should be in general  significantly smaller in Part 3 than that in Part 1.

\begin{figure*}
\hskip  5.0 cm
\centerline{\includegraphics[width=1.1\textwidth]{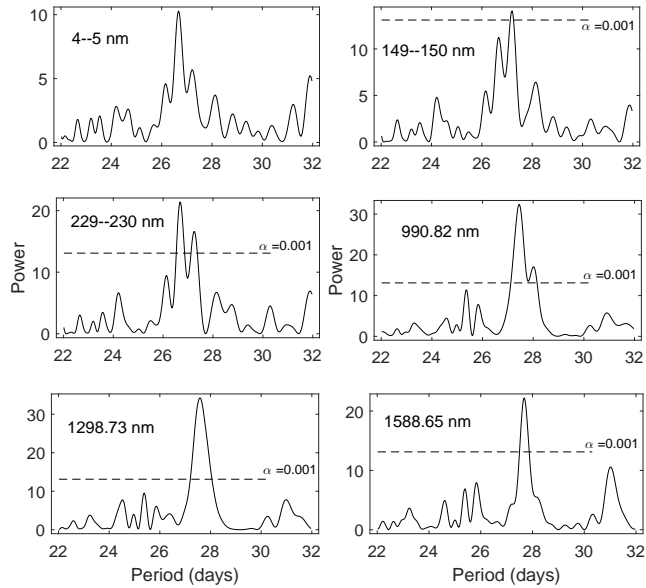}}
\caption{The same as Figure 4, but just SSIs in ascending time of cycle 24 (the years of 2008.958 to  2014.292)  are used to calculate power spectra.
}\label{}
\end{figure*}
\begin{figure*}
\hskip  5.0 cm
\centerline{\includegraphics[width=1.0\textwidth]{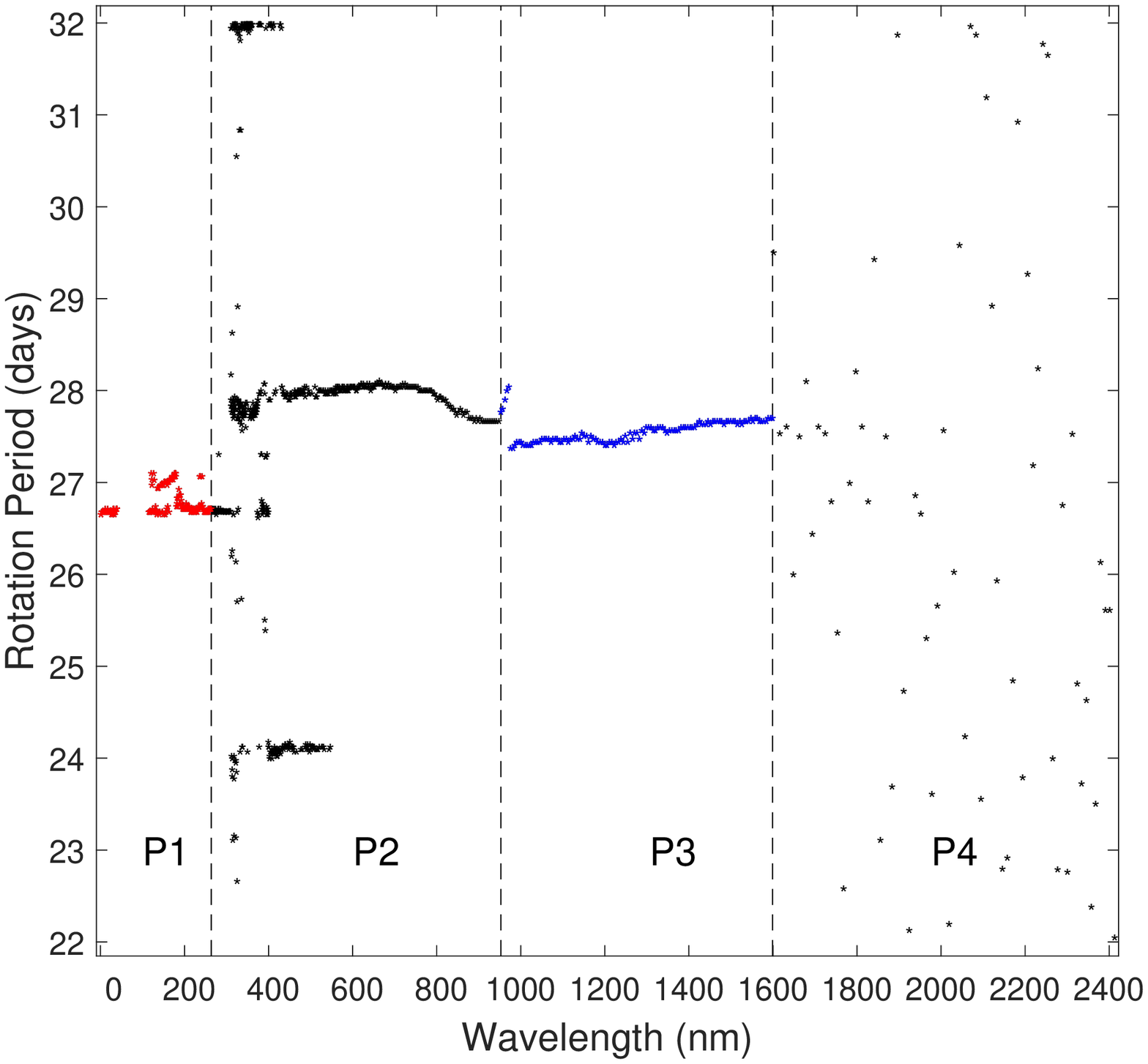}}
\caption{The same as Figure 5, but just SSIs in  ascending time of cycle 24   are used to to calculate power spectra, and then to determine rotation periods.
}\label{}
\end{figure*}
\begin{figure*}
\hskip  5.0 cm
\centerline{\includegraphics[width=1.0\textwidth]{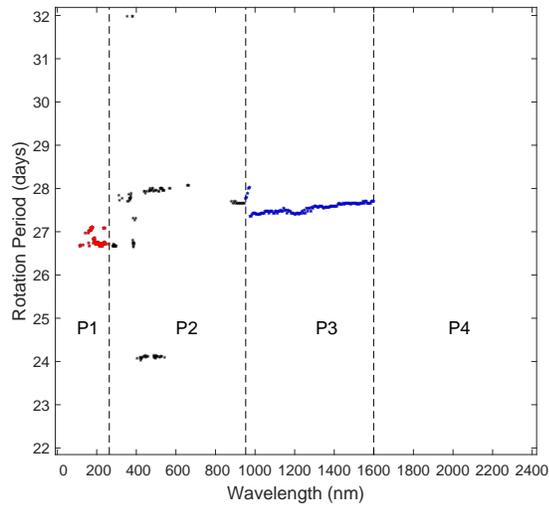}}
\caption{The same as Figure 6, but just SSIs in  ascending time of cycle 24   are used to to calculate power spectra, and then to determine rotation periods.
}\label{}
\end{figure*}

\subsection{Rotation period for the descending time of cycle 24}
Yearly number of no-data days is counted up for each SSI, and then it is shown in Figure 16 just for 6 bands of SSIs. The last two panels in the figure indicate that there are very few data left in the descending time of cycle 24 for long wavelength bands, and Figure 2 also displays big data missing for long wavelength bands. The above analyses indicates irregular variation of rotation periods for long wavelength bands in Part 4. Therefore the SSIs in this part will not be considered. The first 921 SSIs in  the descending time of cycle 24 (from  the year of 2014.292 to October 28 2017) are used to calculate individual power spectra by means of the same periodogram method, and Figure 17 shows the power spectra of the 6 selected bands of SSI  and their corresponding statistical test lines at the $0.1\%$ significance level.  Then
similarly, Figure 18 shows the  period of solar rotation for each SSI while its statistical significance is not considered, and Figure 19, those  periods of solar rotation which are statistically significant at the $0.1\%$ level.
The distribution of synodical rotation period varying with SSI can be divided into the same three parts as Figures 5 and 6 do, and
for Parts 1 and 3, rotation period hardly changes with wavelength.
The relevant values are listed in Table 1, and
the mean rotation period is found to be significantly longer in Part 3 than that in Part 1.  Actually in Figures 18 and 19, all rotation-period  values in Part 1 are obviously smaller than those in Part 3, indicating that rotation period should be in general  significantly longer in Part 3 than that in Part 1.

\begin{figure*}
\hskip  5.0 cm
\centerline{\includegraphics[width=1.7\textwidth]{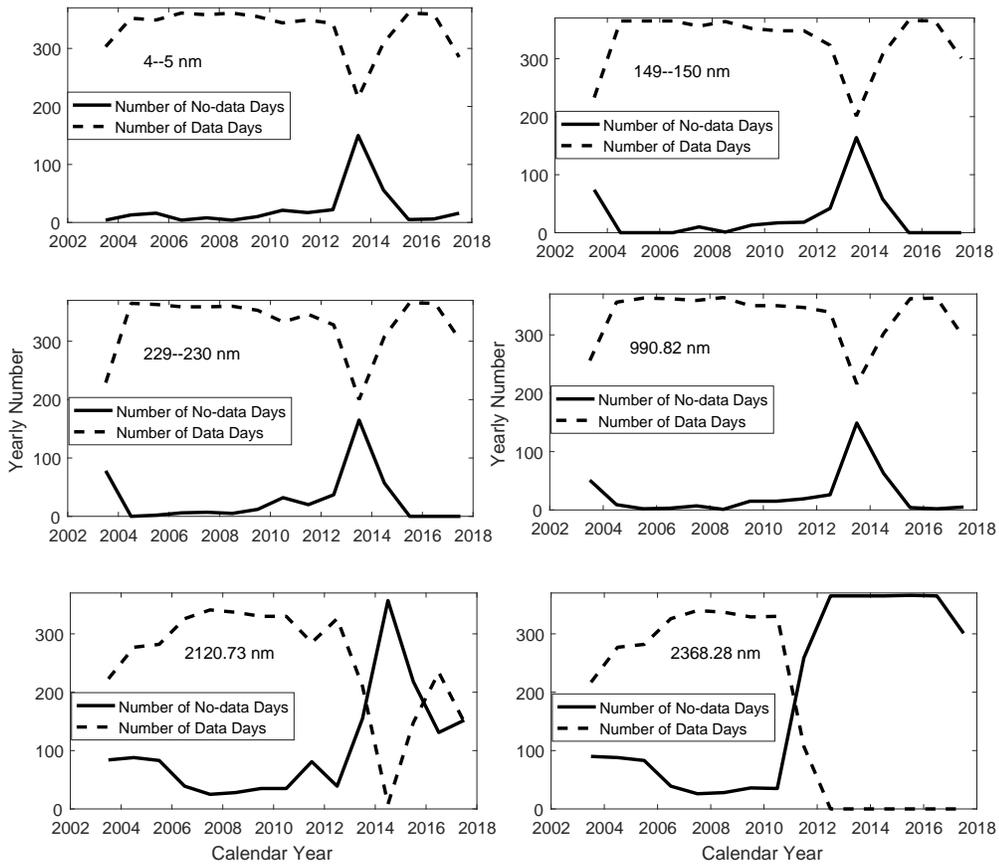}}
\caption{Yearly number of no-data days (solid lines) and that of data days (dashed lines) for 6 SSIs whose wavelengths are shown in each panel.
}\label{}
\end{figure*}
\begin{figure*}
\hskip  5.0 cm
\centerline{\includegraphics[width=1.1\textwidth]{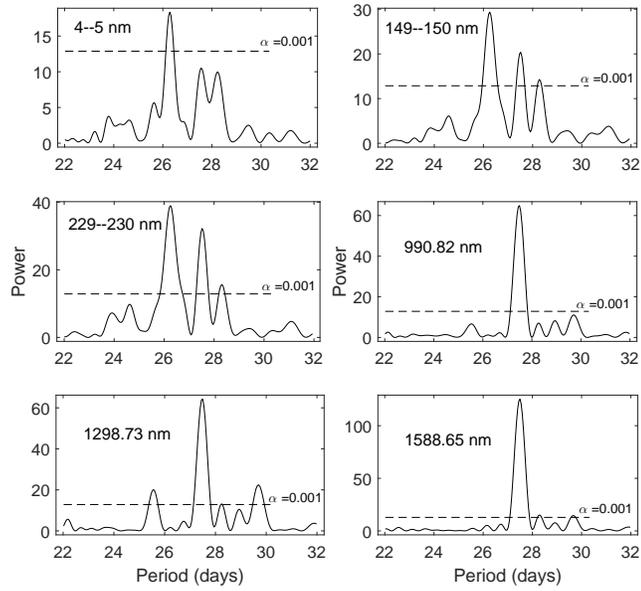}}
\caption{The same as Figure 4, but just SSIs in descending time of cycle 24 (the year of 2014.292 to October 28 2017)  are used to calculate power spectra.
}\label{}
\end{figure*}
\begin{figure*}
\hskip  5.0 cm
\centerline{\includegraphics[width=1.0\textwidth]{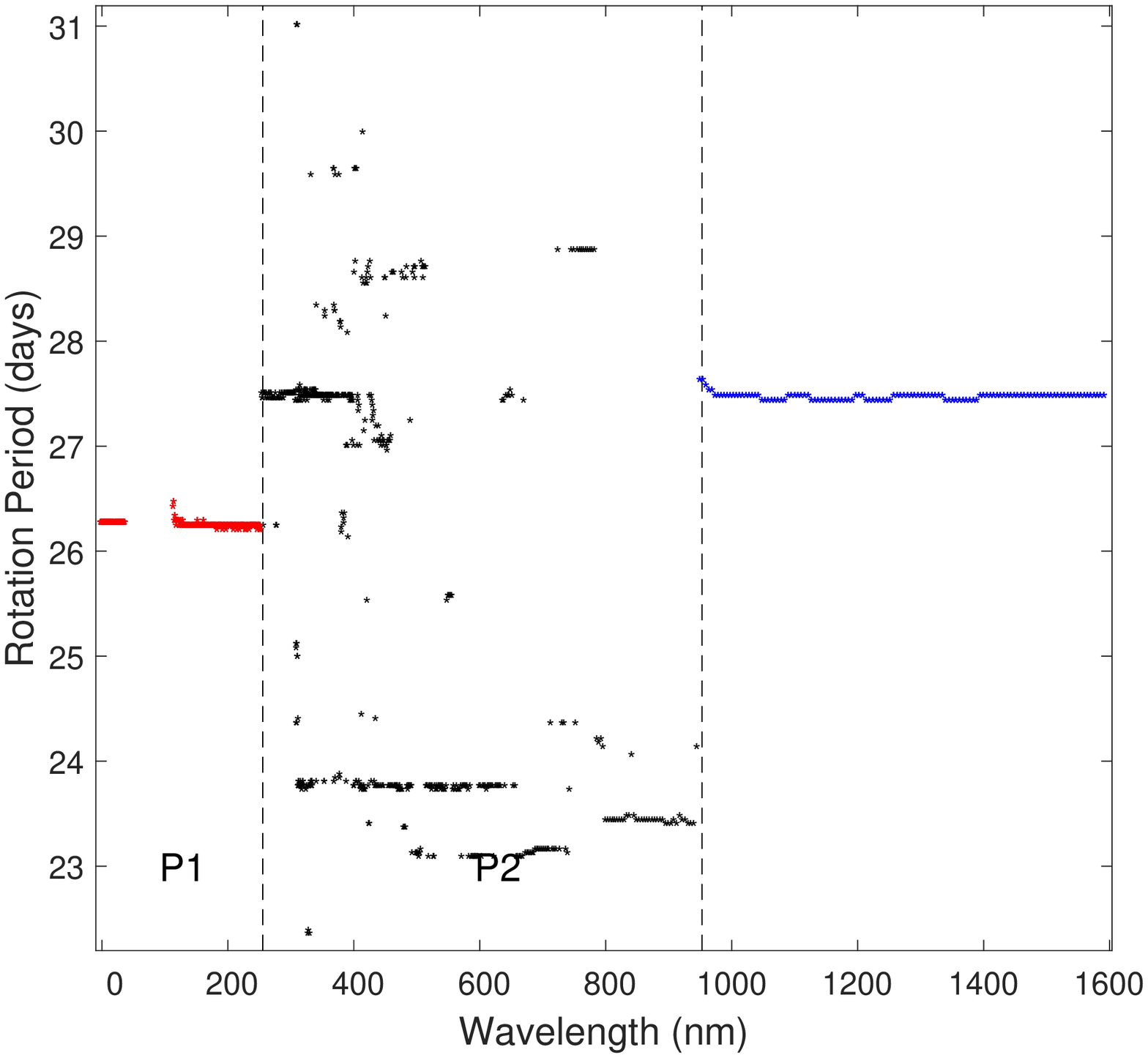}}
\caption{The same as Figure 5, but just SSIs in  descending time of cycle 24   are used to to calculate power spectra, and then to determine rotation periods.
}\label{}
\end{figure*}
\begin{figure*}
\hskip  5.0 cm
\centerline{\includegraphics[width=1.6\textwidth]{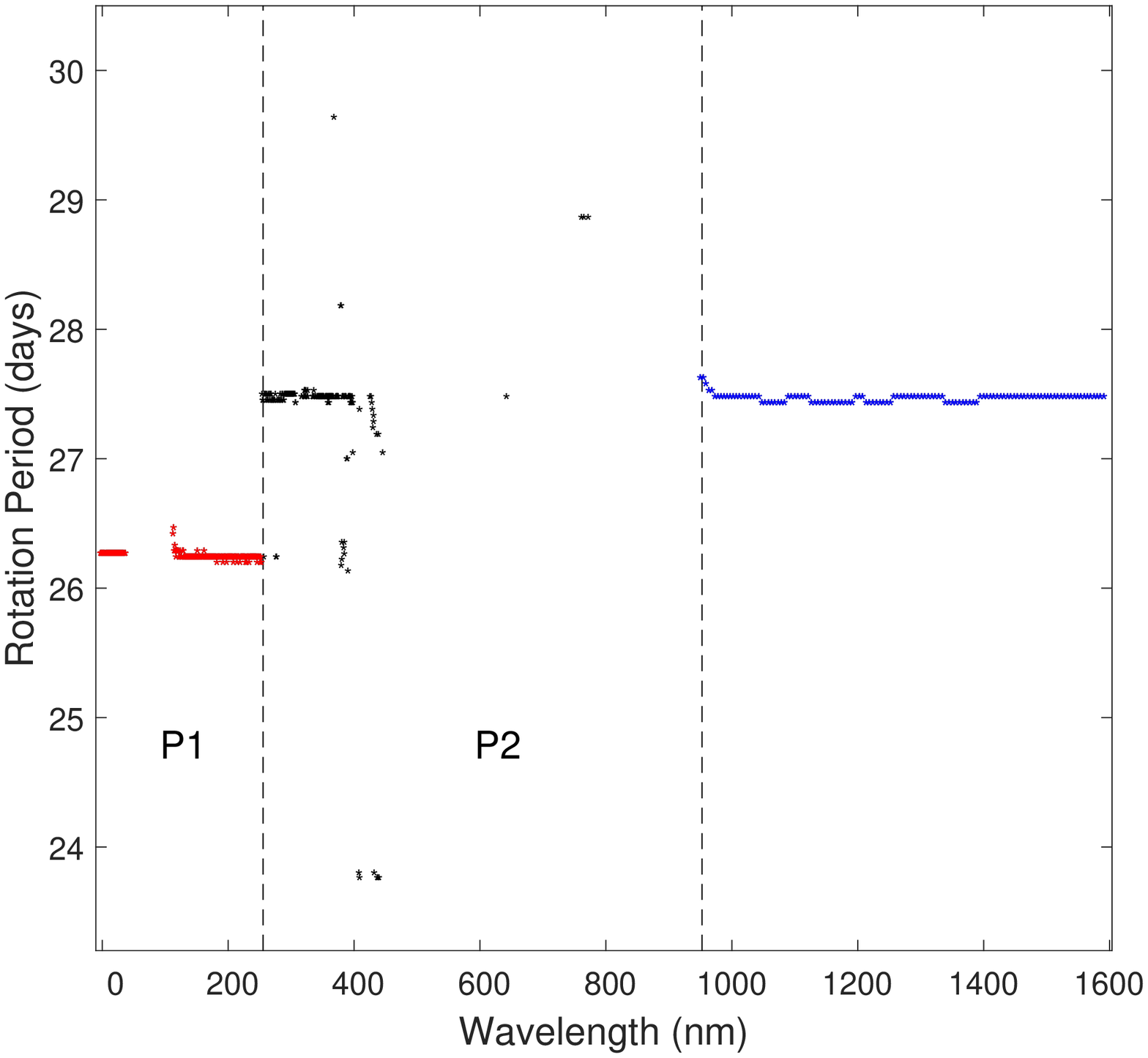}}
\caption{The same as Figure 6, but just SSIs in  descending time of cycle 24   are used to to calculate power spectra, and then to determine rotation periods.
}\label{}
\end{figure*}

\section{Conclusions and discussion}
In this study, 985 SSIs with wavelengths spanning 1 $\sim$ 2400 nm  are utilized to investigate characteristics of synodical rotation period in the solar atmosphere, from the photosphere to  corona.
They are measured during March 1 2003 to October 28 2017 by the SORCE satellite, and
their daily time series are all uneven, and thus the practical Lomb \,-\, Scargle periodogram method is used to calculate their power spectra. Resultantly,
the SSIs whose wavelengths are  shorter than 264 nm  are found to present an approximately constant synodical rotation period, $26.327\pm0.129$ (days), and for the SSIs whose wavelengths are  shorter than 40 nm  rotation period is even a constant. The SSIs with wavelengths between 952 and 1600 nm are also found to display an approximately constant synodical rotation period, $27.545\pm0.055$ (days). Statistical test confirms that the first approximate constant is significantly shorter than the second. For each of the SSIs whose wavelengths are shorter than 40 nm or longer than 952 nm but shorter than 1600 nm, the power spectrum around  rotation period scale ($27.5\pm 5$ days) shows that just one  or no peak is statistically significant.

{  It is known that SSIs whose wavelengths are shorter than 40 nm form in the solar corona, SSIs whose wavelengths are  longer than 40 nm but shorter than 264 nm  form  in the solar corona and transition region, and SSIs whose wavelengths are  greater than 950 nm but less than 1600 nm  form  at the bottom of the photosphere (Ding $\&$ Fang 1989; Harder et al 2009; Meftah et al 2018; Li et al. 2018). That is, X-rays and far and middle ultraviolet spectra mainly form in the transition region and corona, the SSIs in Part 1 (P1) covers wavelengths of $1\sim 264$ nm, and thus the forming height of the SSIs in Part 1 corresponds to the corona and transition region. The SSIs in Part 3 (P3) locate at infrared wavelengths ($900\sim 1600$ nm),  and the forming height of the SSIs in Part 3 corresponds to the low photosphere.}

There are two kinds of rotation in the solar atmosphere, one is rotation of the whole atmosphere plasma, and the other is that of  local magnetic structures.
It is believed that coronal heating is in close relation to the magnetic field, and that magnetic structures mainly shape coronal emission structures with violent spatial inhomogeneity  due to coronal low $\beta$ (De Moortel $\&$ Browning 2015). Observation, theoretical, and statistical studies have blamed small-scale magnetic activity for the heating of the entire coronal atmosphere (Parker 1972, 1988; Ji et al. 2012; Parnell 2012; Arregui 2015; Cargill et al 2015; Klimchuk 2015; Longcope $\&$ Tarr 2015; Peter 2015; Schmelz $\&$ Winebarger 2015; Velli et al 2015; Wilmot-Smith 2015; Zhang et al. 2015; Henriques et al 2016; De Pontieu et al 2018; Li et al 2018), and the solar surface is covered by ubiquitous small-scale  magnetic structures, as if the Sun wraps around a magnetic blanket (Zirin 1988; Wilhelm et al 2007).
It is important to notice the coronal atmosphere has to be heated and to be formed in the first place (De Moortel $\&$ Browning 2015).
The corona is highly structured in the form of  substantial magnetic loops with low $\beta$ values (De Moortel $\&$ Browning 2015) (magnetic field ``leads" plasma). Therefore, the rotation period of 26.327 days should be that of coronal plasma atmosphere modulated by small-scale magnetic activity. That is, coronal plasma has the rotation rate of small-scale magnetic activity due to that the entire corona is heated by it, and that ubiquitous small-scale  magnetic structures guide coronal plasma in the structures.

Large sunspots are dark in the photosphere due to their lower temperature than the background photosphere, and their appearance on the solar disk should decrease both solar spectral  and total irradiances (Li et al 2012;  Li et al 2018), therefore, the synodical rotation period of 27.545 days should be the period of the photosphere modulated by large-scale sunspot structures.

It has been known that small-scale magnetic structures rotate faster than large sunspots, and the rotation rate in the photosphere measured by Doppler displacement, which is one of photosphere plasma,  is slower than that measurement by magnetic structures (Howard 1984; Komm et al 1993; Xiang et al 2014; Xu $\&$ Gao 2016).
This is the reason why the corona atmosphere is found here to rotate faster than the photosphere.
Therefore surprisingly, the coronal atmosphere is not only abnormally hotter than the underlying photosphere, regarded as a big question, but also abnormally rotates faster, seeming also a big question. {  In the corona, $(6\sim 15)\times 10^{7} m$ above the solar photosphere which is observed at radio frequencies of $275\sim 2800 MHz$, the inner corona is found to seemingly rotate slightly faster than the outer corona, and the sidereal rotation period is about $22.5\sim 24.4$ (Vats et al. 2001; Chandra $\&$ Vats 2011; Bhatt et al. 2017), much close to our findings (please see Table 1). The sidereal rotation period of the photospherical magnetic structures is about $26\sim 30$ days (Howard $\&$ Harvey 1970), and accordingly the corona is inferred to also rotate faster than the photosphere. Waldmeier (1957), Cooper $\&$ Billings (1962), Vats et al. (1998), and Mancuso $\&$ Giordano (2011) found that the corona  should  rotate faster than the photospherical magnetic structures, somwhat supporting our findings.}

Small-scale magnetic activity not only  makes the corona hotter, but also makes the corona rotate faster than the underlying photosphere.
In other words, the corona atmosphere rotating faster than the photosphere gives evidence for small-scale magnetic activity to heat the corona.
Such a state of the corona is conducive to perturbing the large-scale magnetic field to produce explosions, becoming a new source of disturbance, and more attention should be paid to it  in the future.

In the top photosphere and chromosphere, SSIs are rotationally modulated by  magnetic activities of different scales, and magnetic structures of different scales rotate at different rates. Therefore for these SSIs, peaks of statistical significance are usually clustered around period scale of 27 days, {  and this is inferred to be the major reason why rotation period is dispersively distributed in Parts 2 and 4, because SSIs at visible-light wavelengths and wavelengths of longer than 1600 nm form in the upper photosphere or the chromosphere.}

Because the SSIs at 985 bands are all measured with the Sun regarded as a point of no spatial resolution, all rotation periods obtained here have no bearing on differential,
although rotation is latitudinally differential in the photosphere and corona. That is, all of the obtained periods may be treated as those which are averaged over latitudes. Rotation period usually changes over time. Here in some years rotation period is of significance, but in some years it is insignificant, and temporal variation of  rotation period in the corona is believed to depend on temporal variation of  rotation period of small-scale magnetic activity. Anyway all periods obtained here may be regarded as one kind of period averaged over the considered time, and  as a whole during the time interval the corona rotates faster than the underlying photosphere. Rotation characteristics are also investigated when the entire time interval is divided into four parts, the descending period (2003.162 $\sim$ 2008.958) of solar cycle 23, cycle 24 (2008.958 to 2017.823), and  the ascending (2008.958 $\sim$ 2014.29167) and descending periods (2014.29167 $\sim$ 2017.823) of cycle 24, and for each time part the corona is also found to rotate faster than the photosphere. \\

{\bf Acknowledgments:}
The authors thank the anonymous referee for careful reading of the manuscript and constructive
comments which improved the original version of the manuscript.
The SORCE Solar Spectral Irradiance (SSI) composite data product is
constructed using measurements from the XPS, SOLSTICE, and SIM instruments, which are combined into merged daily solar spectra over the
spectral intervals (1-39 nm and 116-2416 nm).
This work is supported by the
National Natural Science Foundation of China (11633008 and
11573065)  and the Chinese Academy of Sciences.

\clearpage

\end{document}